\documentclass{PoS}

\def\LL{\left\langle}   % left angle bracket
\def\RR{\right\rangle}  % right angle bracket
\def\PAR#1#2{ {{\partial #1}\over{\partial #2}} }

\newcommand{\BE}{\begin{displaymath}}
\newcommand{\EE}{\end{displaymath}}
\newcommand{\BNE}{\begin{equation}}
\newcommand{\ENE}{\end{equation}}
\newcommand{\BEA}{\begin{eqnarray}}
\newcommand{\EEA}{\nonumber\end{eqnarray}}

\title{Improved method for computing nucleon strangeness}

\ShortTitle{Improved method for computing nucleon strangeness}

\author{\speaker{Walter Freeman} \\
        University of Arizona\\
        E-mail: \email{wfreeman@physics.arizona.edu}}

\author{Doug Toussaint\\
        University of Arizona\\
        E-mail: \email{doug@physics.arizona.edu}}

\abstract{The strange quark content of the nucleon, $\LL N|\bar s s|N \RR - \LL 0|\bar s s|0 \RR$, 
as well as other matrix elements, can be calculated on the lattice by 
examining correlations between the nucleon propagator and the quark condensate. 
The largest contribution to statistical error comes from 
fluctuations in the condensate 
far from the propagation region that contribute only noise. 
We will report on a technique for considering only 
the condensate near the propagation region, significantly reducing the statistical error.}

\FullConference{The XXVIII International Symposium on Lattice Field Theory\\
                 June 14-19,2010\\
                 Villasimius, Sardinia Italy}

\begin{document}

\section{Motivation}

The strange quark content of the nucleon is difficult to measure experimentally but is a quantity of wide interest.
In particular, the interaction cross section between some proposed dark matter candidates (for instance, neutralinos)
and ordinary matter may
have a large contribution from Higgs-like exchange with sea strange quarks in the nucleon\cite{BALTZ06,ELLIS08}.
Specifically, the interest is in the quantity $\LL N| \int\, d^3x\, \bar s s |N\RR - \LL 0| \int\, d^3x\, \bar s s |0\RR$ --
the connected part of the strange quark condensate, integrated over all space, in the nucleon.
Recently this quantity has been calculated on the lattice by several groups \cite{OURPRL, BALI09, OHKI09, YOUNGTHOMAS09, JLQCD08, BALI08} using
a variety of approaches. Several other results were also presented at Lattice 2010 \cite{ENGELHARDT10, TAKEDA10, RAMOS10, JUNG10}.
These calculations all have large statistical error, owing to the inherently noisy nature of evaluating
disconnected diagrams. Our previous calculation using the large MILC library of 2+1 flavor Asqtad staggered fermion configurations \cite{RMP} found  
$\LL N| \bar s s |N\RR = 0.69(7)_{stat}(9)_{sys}$. Reduction of this large statistical error is thus
a priority, and might also make possible better control of systematics; in particular, it may
enable a reduction in the systematic error estimate due to excited state pollution of the propagators.

\section{Partial condensate method for reducing statistical error}

In our previous method \cite{OURPRL}, the matrix element $\LL N | \bar s s | N \RR - \LL 0 | \bar s s | 0 \RR$
is equated to ${{\partial M_N}\over{\partial m_s}}$ via the Feynman-Hellman theorem. This can then be written as the
product of two other partial derivatives $\displaystyle\sum\limits_t$ ${{\partial M_N} \over 
{\partial P(t)}} {{\partial P(t)} \over {\partial m_s}}$, where $P(t)$ is the nucleon propagator
and the sum runs over a range of propagator separations $t$. The minimum distance used in this 
range should be chosen large enough that excited states
have mostly decayed away but small enough that the signal-to-noise ratio is large; the result is
relatively insensitive to the maximum distance because of the rapidly declining signal-to-noise
ratio in $P(t)$. Since $M_N$ is a complicated
function of $P(t)$ defined implicitly by a fitting procedure, the first of these
two partial derivatives can be measured most easily by applying small changes to $P(t)$ and examining
the resulting change in $M_N$.
The second of these can be written, using the Feynman-Hellman theorem (in reverse), as
$\LL P(t) \int \, d^4x \, \bar s s \RR - \LL P(t) \RR \LL \int \, d^4x \, \bar s s \RR$. 
This is expedient, since the whole-lattice condensate $\int \, d^4x \, \bar s s$ has been previously measured by MILC
on all of the lattice configurations, as have the propagators $P(t)$. Thus, this method allows a high-statistics calculation
of the nucleon strangeness with no additional expenditure of computer time.

However, a major contribution to the statistical error in this calculation comes from fluctuations
in the quark condensate that have no physical correlation with the hadron propagator. 
While the correlation between these fluctuations and the propagator averages to zero
in the limit of infinite statistics, with finite statistics they do not, and spurious correlations of this 
sort are a major contributor to statistical error. The time length of the lattice is much greater than the actual region over
which the propagator is measured; for instance, on many of the $a=0.12$ fm ensembles, the
lattice has a temporal extent of $64a$, but the longest two-point function used only has a length of $15a$.

Since there is no physical reason that fluctuations in the quark condensate
far from the propagator should be correlated with it, those fluctuations contribute only
noise and can be discarded without introducing bias; 
in other words, we replace 
\BNE \LL P(t) \int \, d^4x \, \bar s s \RR - \LL P(t) \RR \LL \int \, d^4x \, \bar s s \RR \ENE
with \BNE \LL P(t) \int \, d^3x \, \int_{t_1}^{t_2} dt \, \bar s s \RR - \LL P(t) \RR \LL 
\int \, d^3x \, \int_{t_1}^{t_2} dt \, \bar s s \RR \ENE where $t_1$ and $t_2$ 
are chosen sufficiently far from the propagation region so that they do not affect the final result.

By recalculating the quark condensate and independently storing the value of $\bar s s$ on each
timeslice, it is possible to use only those timeslices which have a meaningful correlation with the
propagator, thus reducing noise and statistical error. It is possible that this reduction in 
error will also lead to greater precision in the determination of the dependence
of $\PAR{M_N}{m_s}$ on the minimum distance used in the fit, allowing for a reduction in the 
estimate for the statistical error from excited states.

\section{New calculations}

This also requires recalculation of the hadron propagators themselves. In order to make
most efficient use of the lattice, MILC uses the average of multiple sources, each located on a different
timeslice, to calculate the propagator; in addition, the forward and backward propagators 
for each source are averaged together. Only this average is stored, but since it contains
contributions from the entire time extent of the lattice, it is correlated with the condensate
on every timeslice. By recomputing the hadron propagators and saving a separate propagator
in each direction for each source timeslice, it is possible to extract only the physically-meaningful
correlations.

In the previous calculation, the fit was mostly controlled by the coarsest ($a=0.12$ fm) data;
this is due both to higher statistics of the coarser lattice ensembles and the extra noise 
from the high-momentum modes present in the finer lattices. 
Since it is significantly more expensive to recompute the nucleon propagators
in this manner on finer lattices, and since the coarser lattices dominate the fit in any case,
we have only done this procedure on the $a=0.12$ fm lattices. Four of the five
ensembles have been completed, and a fifth is in progress. 

\section{Validity checking}

\begin{figure}
\vspace{-1.1in}
\begin{center}
\hspace{-0.5in}
\includegraphics[width=3.4in]{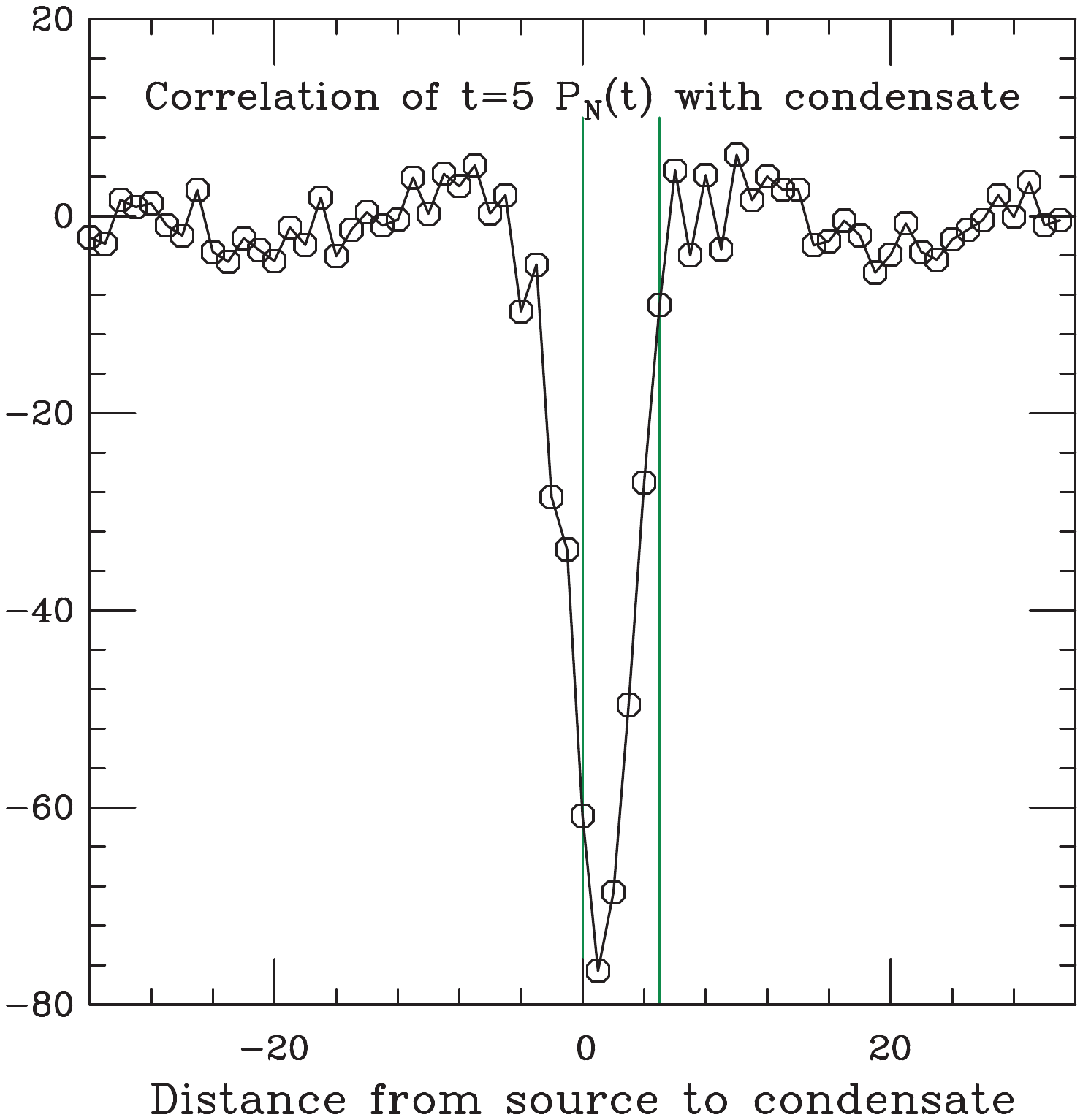}\hspace{-0.5in}
\hspace{-0.5in}
\includegraphics[width=3.4in]{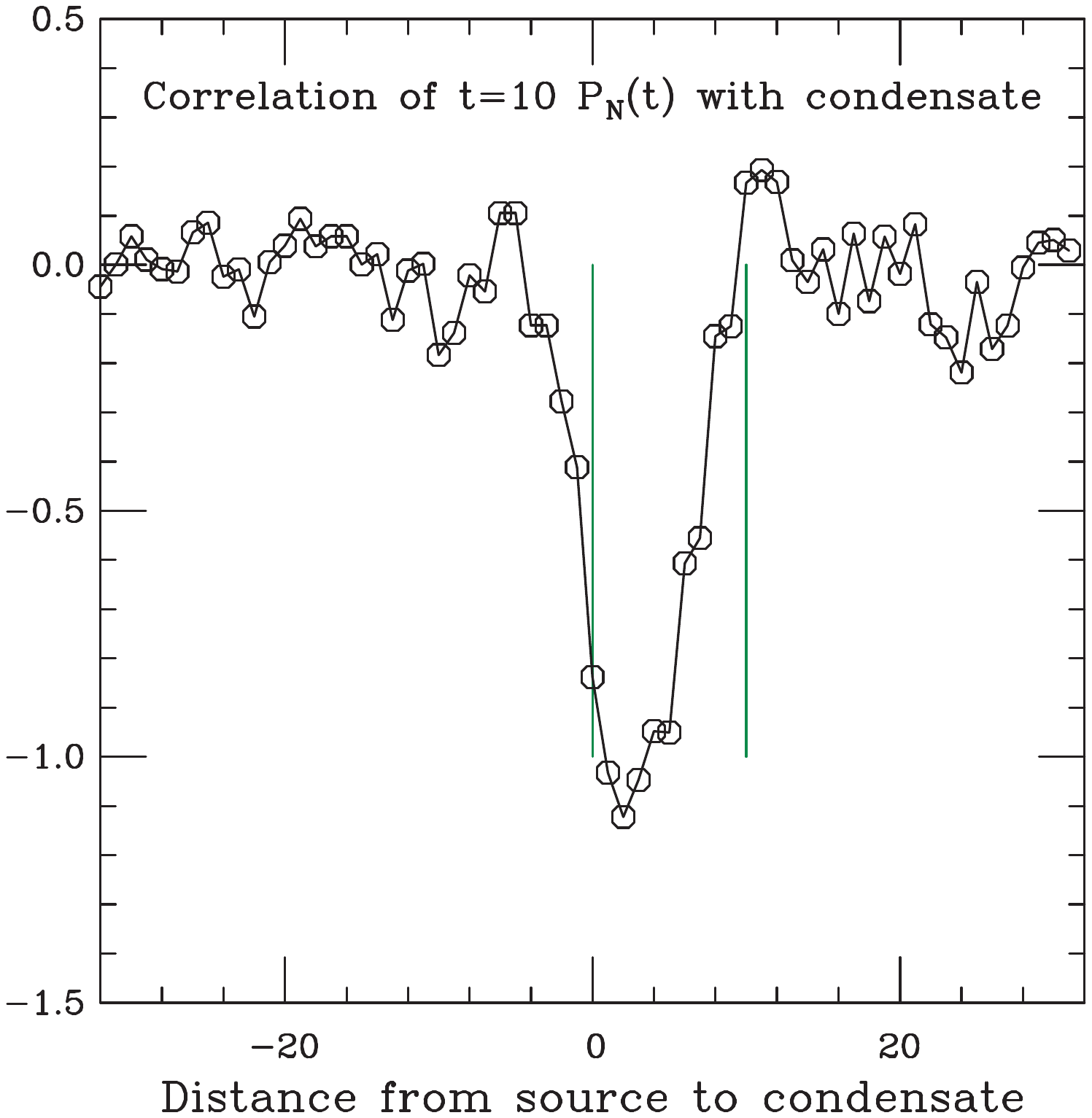}\vspace{-0.5in}
\end{center}
\caption{The correlation between nucleon propagators of length $5a$ and $10a$ 
with the strange quark condensate on different timeslices on
one of the MILC $a=0.12$ fm ensembles. The source and sink of the propagators are marked
in green.}
\label{fig:sbs_prop_corr}
\end{figure}

A larger reduction in statistical error is possible if a smaller range of condensate around 
the propagator is kept, but if too much of the condensate is discarded, it is possible that
some physically-meaningful correlations will be eliminated along with the noise. It is instructive
to look at the correlation between $\int \, d^3x \, \bar s s(t)$ and a propagator with source at $t=0$ 
and sink at $t=T$; we expect this correlation to fall off rapidly outside the region of propagation from
0 to $T$. Figure~\ref{fig:sbs_prop_corr} shows the correlation between this propagator and the strange
quark condensate on different timeslices.

The most notable feature of these plots is the fact that the correlation between
the nucleon propagator and the quark condensate is only distinguishable from zero
in a narrow range around the propagation region for the two short-distance 
propagators. 

\begin{figure}
\vspace{-0.5in}
\begin{center}
\hspace{-0.6in}
\includegraphics[width=0.45 \textwidth]{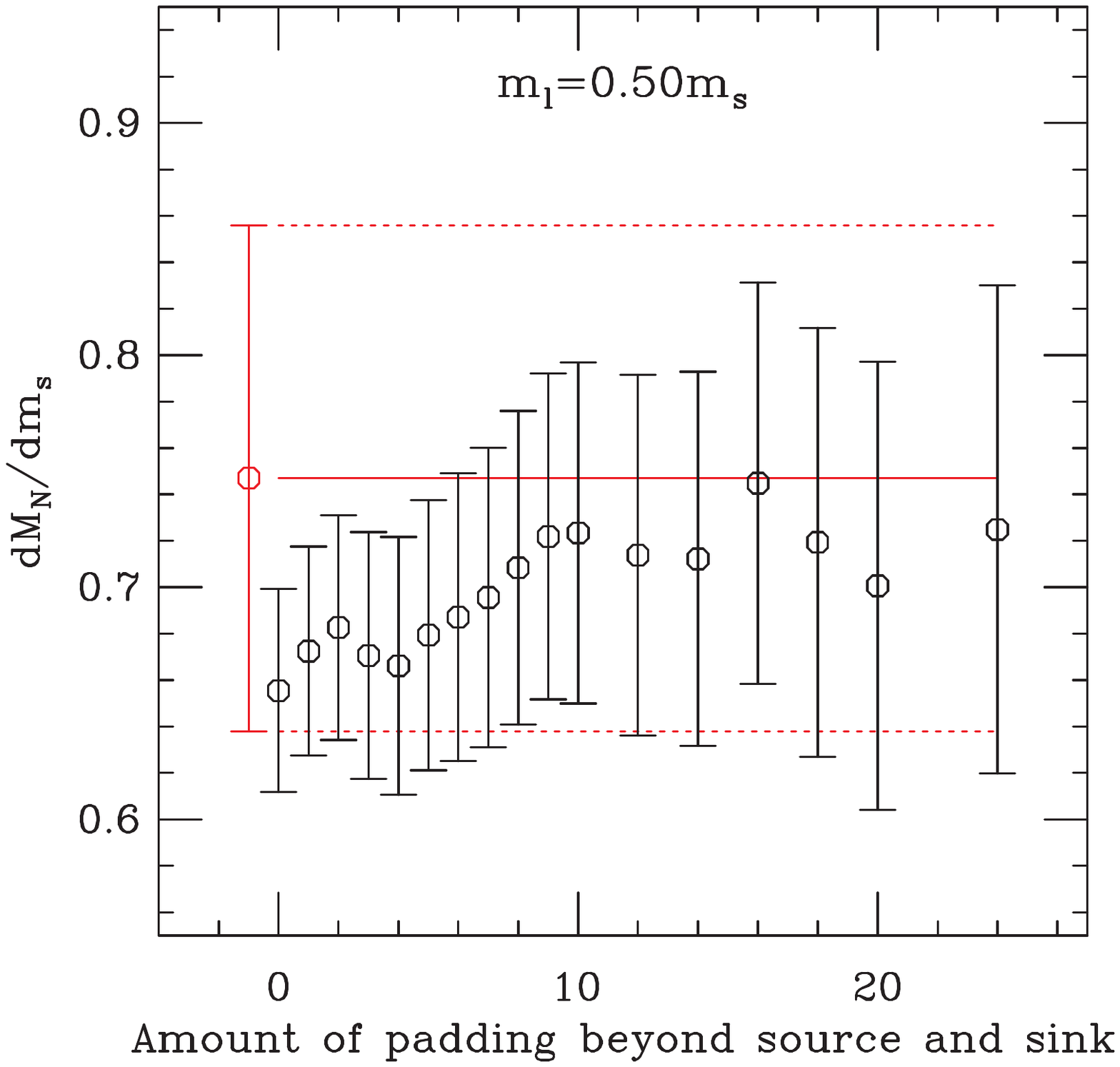}
\hspace{-0.6in}
\includegraphics[width=0.45 \textwidth]{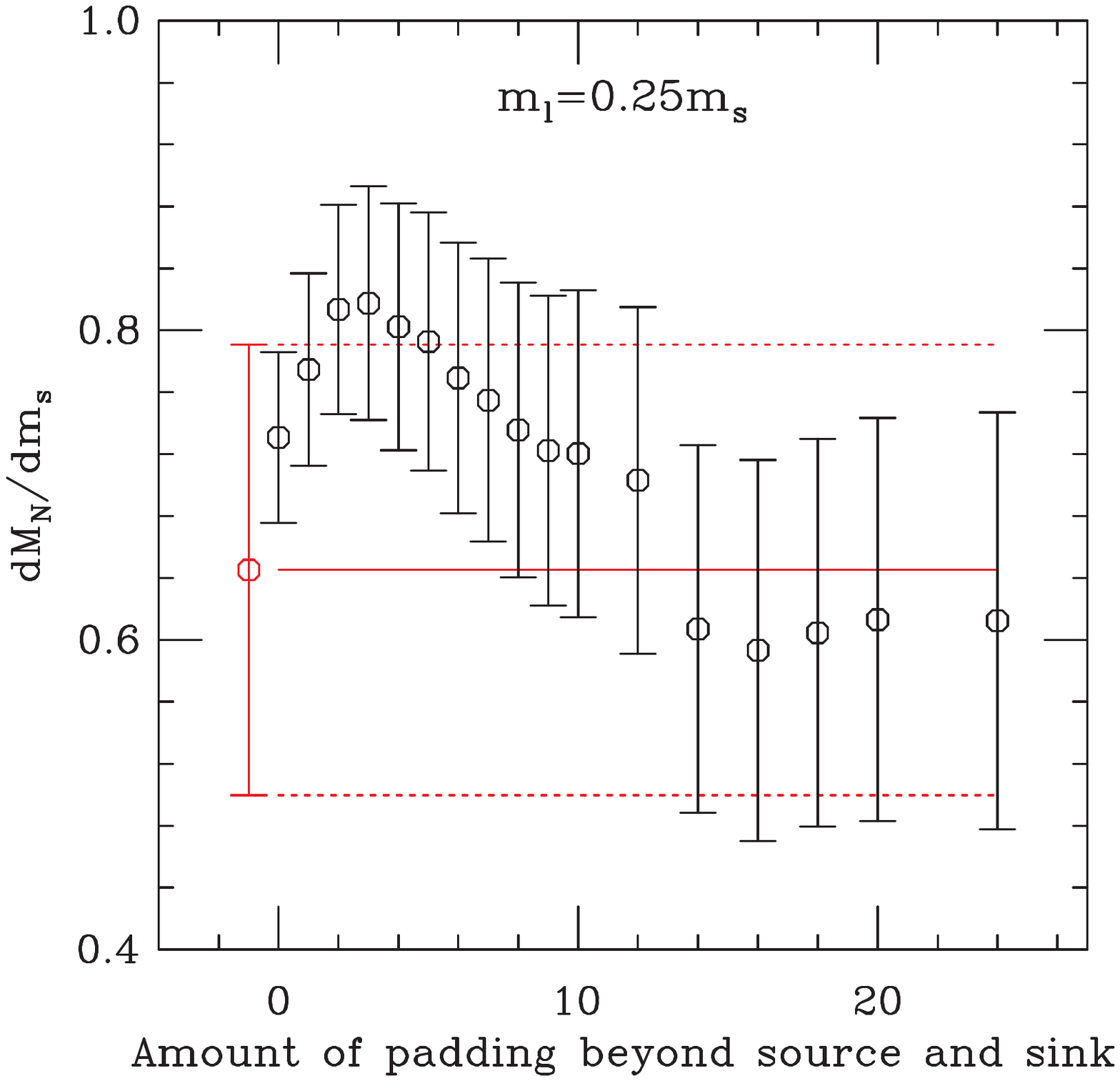} \\
\vspace{-0.9in}
\hspace{-0.6in}
\includegraphics[width=0.45 \textwidth]{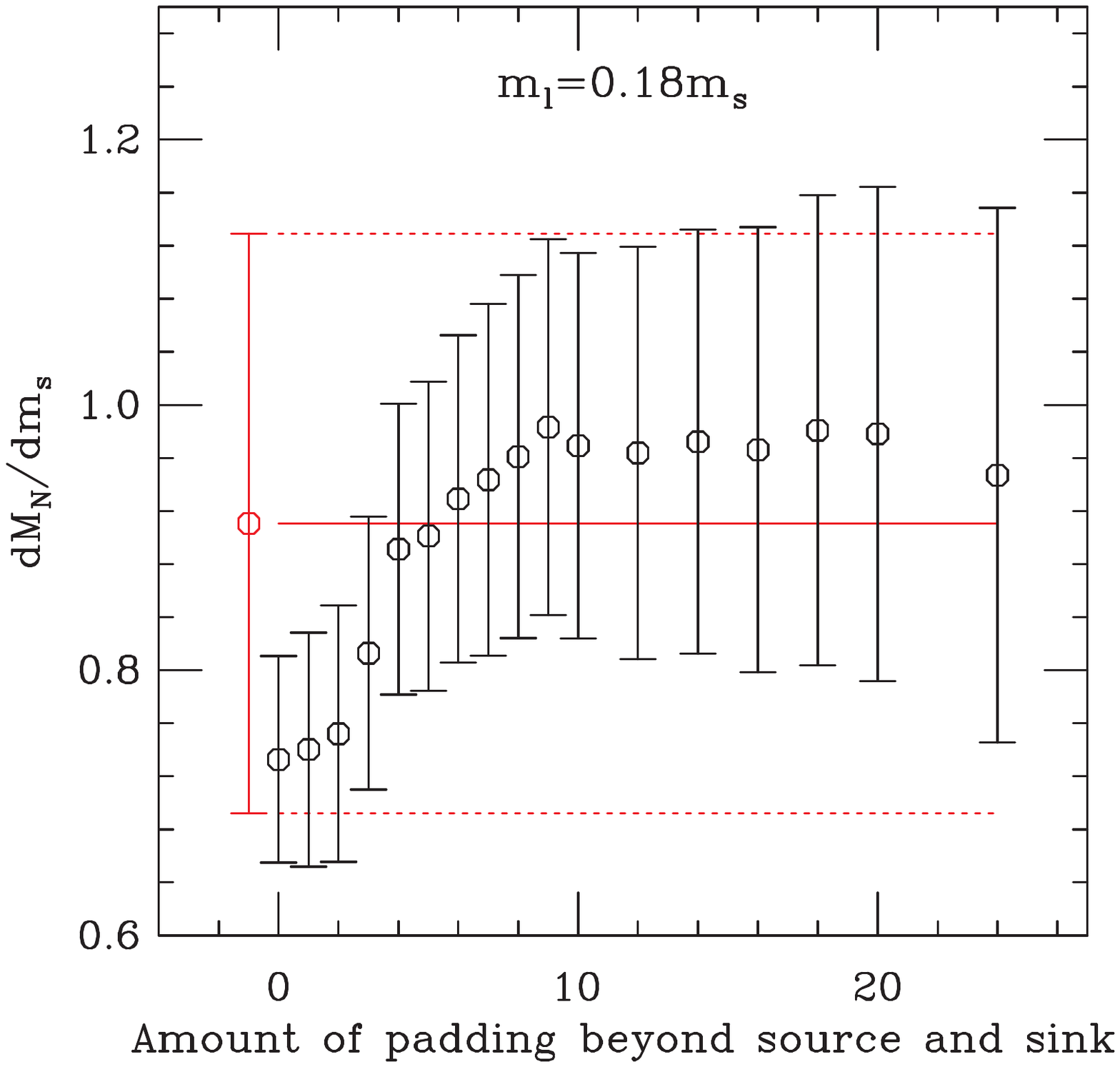}
\hspace{-0.6in}
\includegraphics[width=0.45 \textwidth]{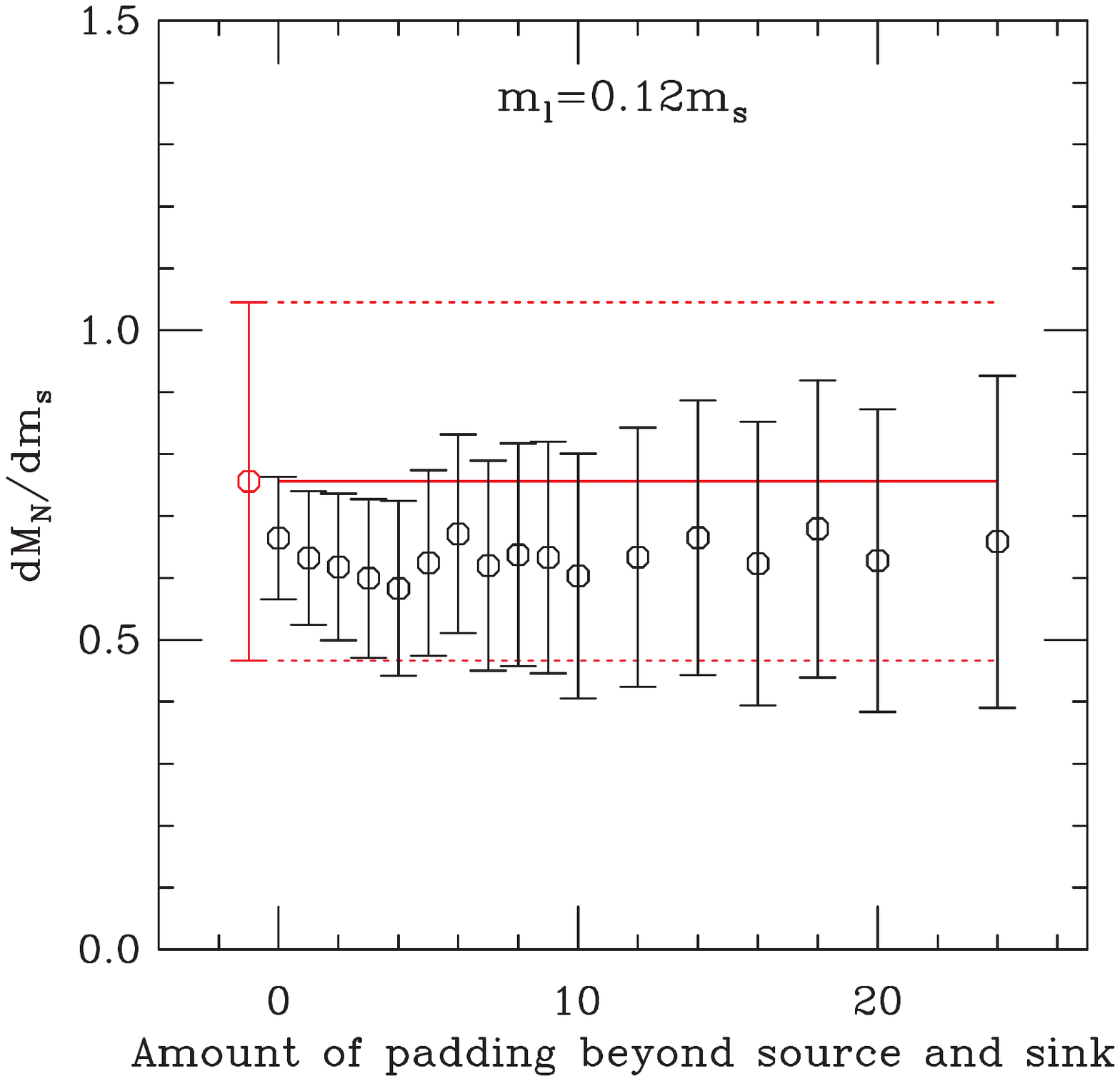}
\end{center}
\vspace{-0.5in}
\caption{The effect of varying the padding size for condensate measurements
on $\PAR{M_N}{m_s}$ for each of the four $a=0.12$ fm ensembles studied.
A padding size of $4a$, for instance, means that the 
condensate is measured in the region between propagator source and sink, and 
for four time units on each end. The red value and dotted lines show the 
result using the old method, in which the condensate is summed over the 
whole lattice.}

\label{fig:padsize_effect}
\end{figure}

These data suggest that it is reasonable to exclude condensate measurements more than
a certain padding distance $T_{p}$ from the propagation region; they suggest that this
distance is in the vicinity of $4-5a$. However, in order to truly understand the effect
that discarding much of the quark condensate has on the result, it is necessary to examine
the dependence of $\PAR{M_N}{m_s}$ itself on the padding size. This dependence is shown in 
Figure~\ref{fig:padsize_effect}.

\begin{figure}
\vspace{-0.85in}
\begin{center}

\includegraphics[width=0.5\textwidth]{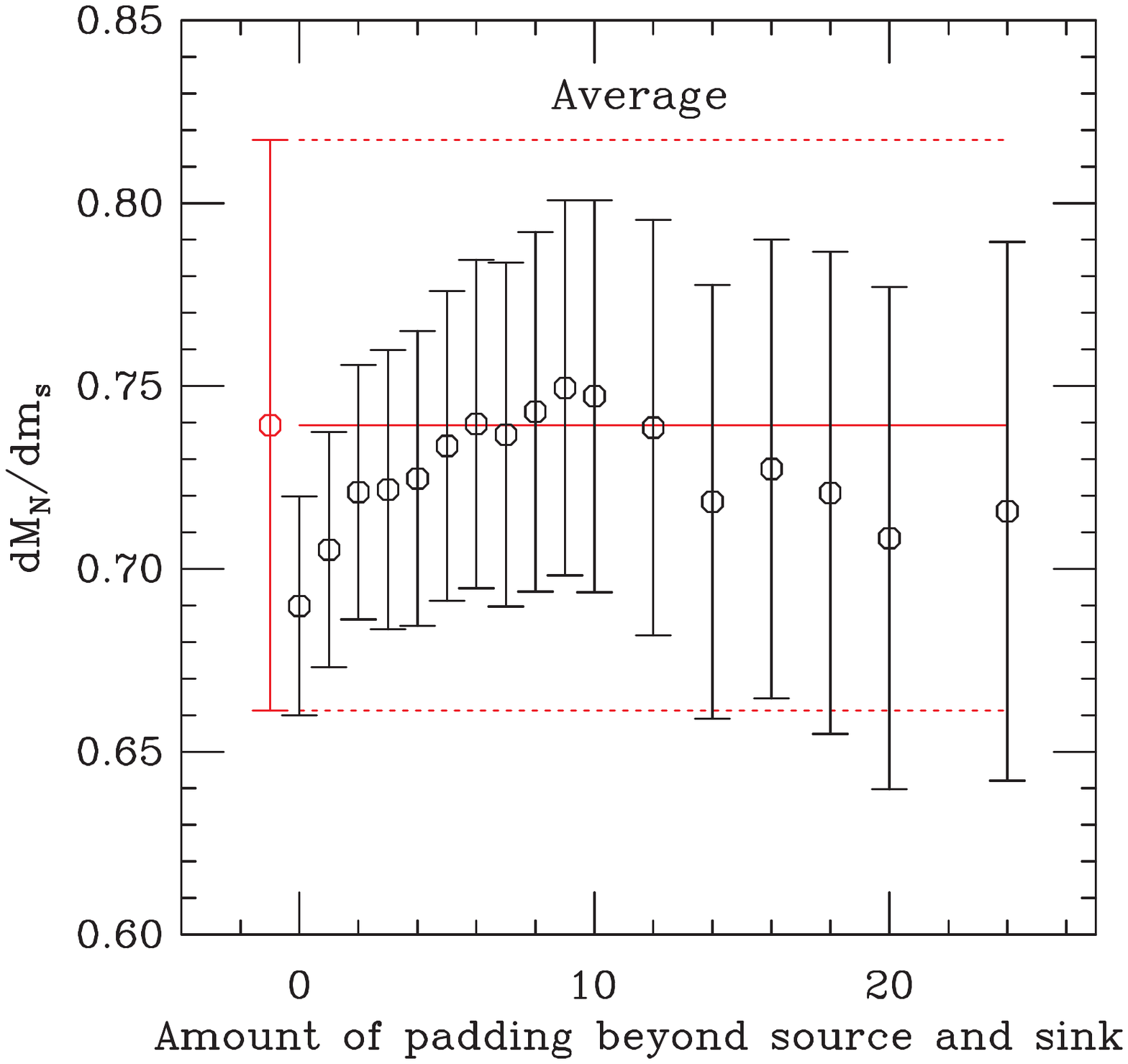}
\end{center}
\vspace{-0.85in}
\caption{The effect of varying the padding size for condensate measurements
on $\PAR{M_N}{m_s}$, averaged over the four $a=0.12$ fm ensembles.
}

\label{fig:padsize_effect_avg}
\end{figure}

Note that results using similar padding sizes are highly correlated, since they are
based on almost the same data set. The most significant effect of this method is the
reduction in statistical error by almost half when using small padding sizes, as expected. 
While most
of the fluctuations in the central value seen here are consistent with statistical effects,
for very small pad sizes a sharp downward trend is noticeable in the results. Since using 
a pad size that is too small amounts to discarding physically significant correlations,
the depressed result for low pad sizes is consistent with the introduction of bias. 

From examining the average of all four ensembles, shown in Figure~\ref{fig:padsize_effect_avg},
it is reasonable to conclude that using a pad size of $4a$ 
introduces no obvious bias and greatly reduces the statistical error. The shift in the central
value is three percent, consistent with the improvement in the statistical error.
Thus it is very unlikely that the systematic error introduced by this method
is larger than that; a more realistic upper bound on the systematic error from this procedure
is one percent.

\section {Effect on systematic error estimates}

\begin{figure}
\vspace{-0.5in}
\begin{center}
\hspace{-0.5in}
\includegraphics[width=0.34\textwidth]{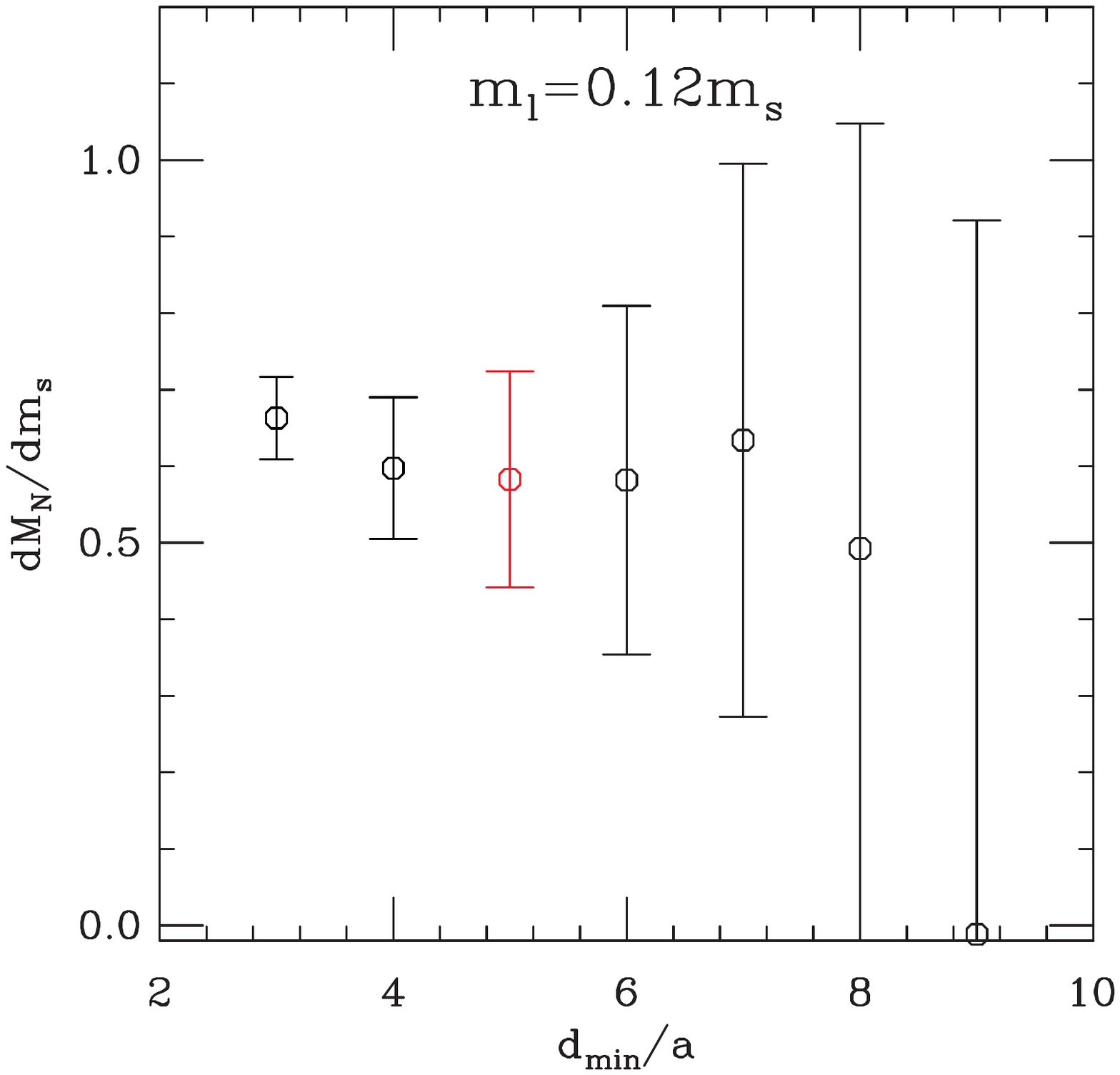}
\hspace{-0.5in}
\includegraphics[width=0.34\textwidth]{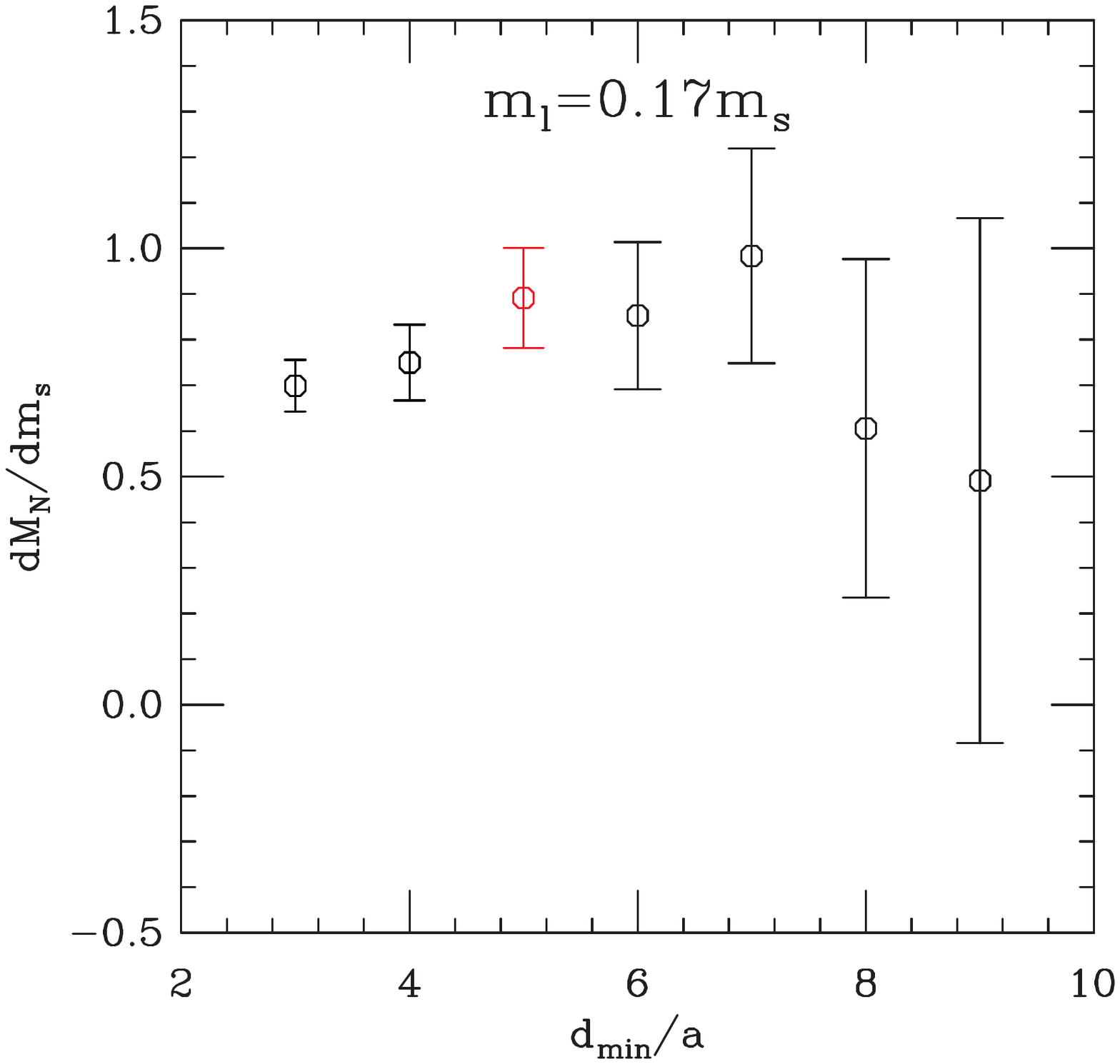} \\
\vspace{-0.9in}
\hspace{-0.5in}
\includegraphics[width=0.34\textwidth]{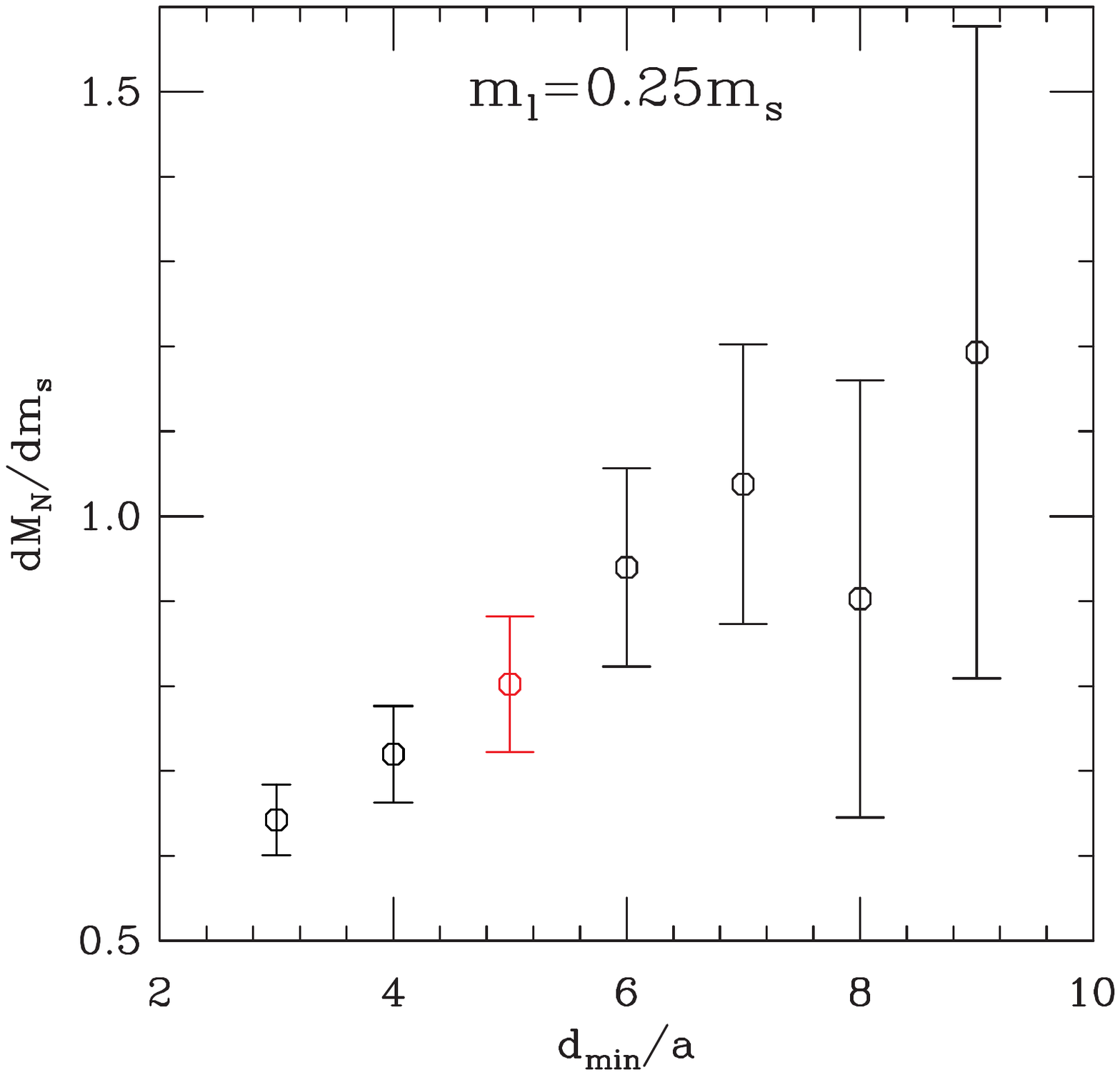}
\hspace{-0.5in}
\includegraphics[width=0.34\textwidth]{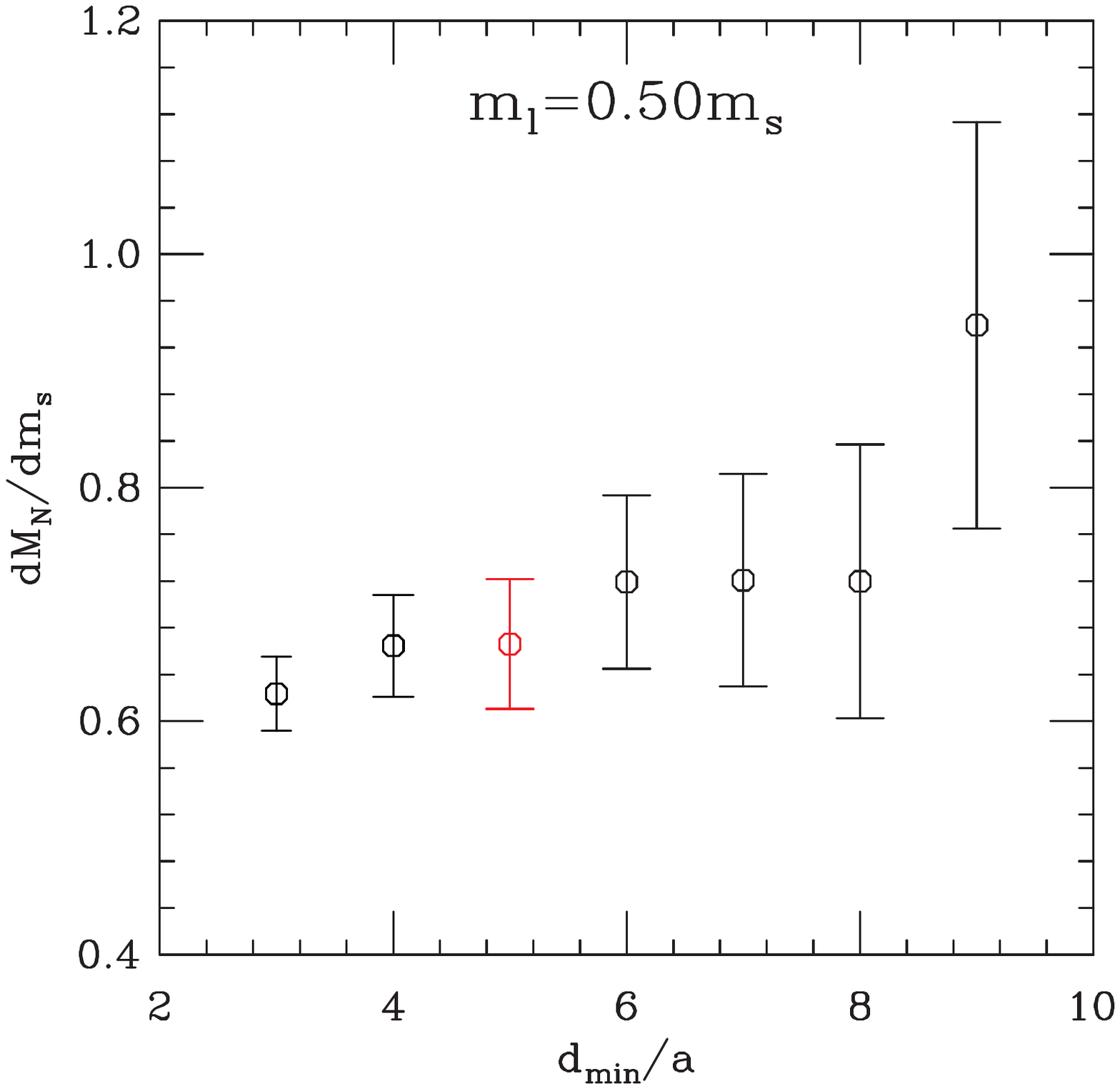}
\end{center}
\vspace{-0.5in}
\caption{The dependence of $\PAR{M_N}{m_s}$ on $d_{min}$ for each of the four ensembles studied.
The red point at $d_{min}=5$ is the point chosen as the best balance between
statistical and systematic error.}

\label{fig:dmin_effect}
\end{figure}

The improved statistical error from this procedure should make improved estimates of the 
systematic error from various sources possible. In particular, a large contributor to the
overall error budget of the previous calculation is the systematic error from excited
state pollution (or, equivalently, the dependence of the result on the minimum fit distance
used). In the previous calculation the statistical errors were so large for values of $d_{min}$
larger than the chosen one that it is difficult to judge whether the dependence of $\PAR{M_N}{m_s}$
on $d_{min}$ is due to systematic or statistical effects. In Ref.~\cite{OURPRL} we chose to give a
conservative estimate for this systematic error; with the better statistics from this method it 
may be possible to reduce this estimate. Here we repeat the analysis of the 
dependence of $\PAR{M_N}{m_s}$ on $d_{min}$ using the new method with
a pad size of $4a$. The results are shown in Figure~\ref{fig:dmin_effect}.

Three of these four ensembles, along with preliminary data from a fifth,
show no strong dependence of $\PAR{M_N}{m_s}$ on $d_{min}$, 
but the $m_l = 0.2 m_s$ ensemble shows a significant effect. It is likely that
this effect is simply a statistical fluctuation rather than an indication that our
value of $d_{min}$ creates a large systematic error, but it prevents us from confidently
giving a lower estimate of the systematic error from excited states. The fifth $a=0.12$ fm 
ensemble, when completed, may enable a lower estimate.

Using this procedure to improve the statistics on the finer lattice ensembles may help to
improve the continuum extrapolation, which was rather poorly constrained in our previous
calculation. However, the effect on the overall error budget from this will be minimal
for the large amount of computer time required to redo nucleon propagators on the $a=0.09 fm$ 
and/or $a=0.06 fm$ lattices.

\section{Conclusion}
By considering only the part of the strange quark condensate that has physically relevant
correlations with the nucleon propagator, the noise contributed by fluctuations in the condensate
far from the propagator can be eliminated and the statistical error greatly reduced. In the 
four ensembles examined here, considering only the condensate within $4a$ of the 
propagation region reduces the statistical error by nearly half without introducing any obvious
systematic bias. The shift in the central value by $3\%$ is explained simply by the reduction in
statistical error, and these results are consistent with the previous MILC result \cite{OURPRL}.

The new data computed for this method (independent nucleon propagators for every source timeslice,
and values of $\bar s s$ on every timeslice) can also be used for a direct evaluation of the 
three-point function $\LL N| \bar s s |N\RR - \LL 0| \bar s s |0\RR$. Preliminary results from this method
are consistent with results obtained with the method presented here. 

A new calculation of the nucleon strangeness using the improved values of $\PAR{M_N}{m_s}$ on the $a=0.12$ ensembles presented here 
is in progress.

\end{document}